\documentclass[sigconf]{acmart}

\copyrightyear{2026}
\acmYear{2026}
\setcopyright{cc}
\setcctype{by}
\acmConference[CHASE'26]{International Workshop on Cooperative and Human Aspects of Software }{April 12--18, 2026}{Rio de Janeiro, Brazil}
\acmBooktitle{International Workshop on Cooperative and Human Aspects of Software (CHASE'26), April 12--18, 2026, Rio de Janeiro, Brazil}
\acmPrice{}
\acmDOI{10.1145/3794860.3794913}
\acmISBN{979-8-4007-2484-8/2026/04}

\usepackage{multirow}
\usepackage{makecell}
\usepackage{tabularx}

\begin{document}

\title{Role and Identity Work of Software Engineering Professionals in the Generative AI Era}

\author{Jorge Melegati}
\orcid{https://orcid.org/0000-0003-1303-4173}
\email{melegati@fe.up.pt}
\affiliation{%
  \department{INESC TEC, Faculty of Engineering}
  \institution{University of Porto}
  \city{Porto}
  \country{Portugal}
}


\begin{abstract}
  The adoption of Generative AI (GenAI) suggests major changes for software engineering, including technical aspects but also human aspects of the professionals involved. One of these aspects is how individuals perceive themselves regarding their work, i.e., their work identity, and the processes they perform to form, adapt and reject these identities, i.e., identity work. Existent studies provide evidence of such identity work of software professionals triggered by the adoption of GenAI, however they do not consider differences among diverse roles, such as developers and testers. In this paper, we argue the need for considering the role as a factor defining the identity work of software professionals. To support our claim, we review some studies regarding different roles and also recent studies on how to adopt GenAI in software engineering. Then, we propose a research agenda to better understand how the role influences identity work of software professionals triggered by the adoption of GenAI, and, based on that, to propose new artifacts to support this adoption. We also discuss the potential implications for practice of the results to be obtained.  
\end{abstract}




\maketitle

\section{Introduction}

The use of Generative AI (GenAI) evokes a disruption in software engineering practice, promising to increase developers' productivity and to improve the quality of produced software~\cite{Ebert2023,Fan2023}. This possibility led researchers to explore the potential implications of GenAI for the human aspects of software engineering~\cite{Malheiros2024,Melegati2024,Schmitt2025,Rauf2025}. Among these aspects, there are consequences to how developers perceive themselves in work~\cite{Melegati2024,Schmitt2025}. In literature, several related concepts are associated with this concern. In general terms, work identities are ``self-meanings tied to participation in work-related activities, such as organizational, occupational and role identities''~\cite{Caza2018}. Related to a specific profession, such as as software engineer or developer, it is relevant to consider the concept of professional identity, which is the constellation of attributes, skills, functions, beliefs, values, motives, and experiences in terms of which people define themselves in a professional role~\cite{Ibarra1999,Fitzgerald2020}. For example, in a seminal paper, Pratt et al.~\cite{Pratt2006} investigated the construction of the professional identity of medical residents. The processes involved in shaping work identities, such as constructing, revising and rejecting~\cite{Caza2018}, are referred to as identity work~\cite{Watson2008}. Research has observed that mismatches between the work and the identity are triggers to identity work~\cite{Pratt2006}. 

Schmitt et al.~\cite{Schmitt2025} presented initial results on the identity work of software engineers with GenAI adoption. Based on 11 interviews with software engineers from a company, the authors observed an influence of the professional seniority on the identity work. However, besides being restricted to few interviews in a single organization, the study did not analyze the effect on identity work of the software engineers' roles. In this paper, we argue that this characteristic of software professionals might be a determinant of identity work triggered by the adoption of GenAI. For example, software developers and testers might perform different changes to their work identities to adapt to this new context. Based on this vision, we propose a research agenda to guide future research in this topic. Better understanding these implications are essential given the importance of self-perceptions for motivation~\cite{Gecas1991,Rosso2010}.

This paper is organized as follow. In Section~\ref{sec:heterogeneity}, we present literature to support that software engineering professionals perceive themselves as an heterogeneous group depending on specific tasks they perform. Then, in Section~\ref{sec:gen_ai}, we discuss how GenAI can change software engineering and the roles involved, identifying the research problem. Section~\ref{sec:agenda} presents a research agenda to tackle this problem. Section~\ref{sec:related_work} reviews related work, including some of the theories that could support the research and existent results on human aspects of software engineering related to work identities. Finally, Section~\ref{sec:conclusions} concludes the paper.

\section{Heterogeneity of software engineering professionals}
\label{sec:heterogeneity}

Despite increased cross-functional skills, especially in the context of agile software development~\cite{Hoda2013}, the software engineering profession comprises several roles, such as developers, testers and architects, as, for example, job advertisements show~\cite{Florea2023}. Even though some research in human aspects of software engineering still make no distinction among different roles, some results show that different roles are perceived diversely, influencing how the professionals might feel. An evident example of this distinction is the role of software architect. These professionals have been seen as technical leaders, defining not only the architecture but also leading and mentoring the team and managing the project~\cite{Eeles2006,Kruchten2008,Sherman2015}. Nevertheless, they are not interested in sharing architectural knowledge, even with the support of automatic tools, suggesting a preference for being ``lonesome decision-makers''~\cite{Hoorn2011}. Therefore, it is expected that they will probably have a different professional identity from their colleagues in development teams.

There is also evidence of difference in perception on the identities of ``non-leading'' roles, namely between software testers and developers. Waychal et al.~\cite{Waychal2021} conducted a survey with 220 software professionals from four countries, i.e., India, Canada, Cuba and China, to identify the advantages and disadvantages of a career on software testing from the perspective of professionals. Among the advantages, they identified learning opportunities, the importance, ease, and availability of the job. But, among the disadvantages, they observed that a major factor was a perception of the role as ``a second-class citizen'', since they are not ``involved in decision making'', are ``blamed for poor quality'', and ``struggle for recognition''. Besides that, the work is viewed as complex, stressful, frustrating, tedious, less creative and not challenging. They concluded that very few professionals are attracted to a career on software testing. Capretz et al.~\cite{Capretz2015} reached a similar conclusion in a study linking personality types and preferred roles in software development. They concluded that software testing is one of the least sought careers in software development across different personality types~\cite{Capretz2015}. In a survey on what developers think about testing, Straubinger and Fraser~\cite{Straubinger2023} observed that software testing is considered a ``mundane'' task and developers prefer to focus on other tasks. Research has shown that this difference affects individual aspects of software testers, such as the perception of their own abilities compared to developers~\cite{Zhang2018}, and the emergence of conflicts between developers and testers~\cite{Cohen2004}.

In summary, rather than a homogeneous group, evidence suggests that software professionals perceive differently themselves and their colleagues based on their role, regarding not only tasks but also prestige. Besides that, even without being formalized, these perceptions mold a hierarchy of roles, even between software developers and testers, which, in principle, could be considered at the same level. Given these differences, a relevant question regards if, and how, these different roles perceive relatedness to each other. In other words, do they identify themselves as being part of a same profession or different professions? In the following section, we will argue that the adoption of GenAI by software engineering teams represents a possible disruptive factor for this scenario. 

\section{GenAI as a disrupting factor for software engineering}
\label{sec:gen_ai}

Since its emergence, GenAI has been extensively explored to support software engineering, both in the industry, as shown by commercial tools, such as GitHub Copilot\footnote{https://github.com/features/copilot} and Cursor\footnote{https://cursor.com/}, and academia, e.g., ~\cite{Liang2024,Davila2024,Sergeyuk2025}. However, the use of GenAI for software engineering creates new risks given its lack of determinism and explainability~\cite{Ebert2023,Fan2023}. For example, in a survey with developers, the most important reasons given to not use AI assistants were not generating code addressing certain requirements and the lack of control of the tool to generate the desired output~\cite{Liang2024}.

To mitigate these issues, an approach that has been explored is the Assured LLM-based Software engineering~\cite{Alshahwan2024}. It consists of the application of LLMs response to software engineering in which all LLM responses come ``with some verifiable claim to its utility''~\cite{Alshahwan2024}. This idea is inspired, in general, by search-based software engineering (SBSE), which has researched ways ``to achieve robust scientific evaluation in the presence of noisy, non-deterministic, and incomplete results''~\cite{Fan2023}, and, in particular, by a generate-and-test approach used by genetic improvement~\cite{LeGoues2019}.

An example of this approach can be seen in the exploration of the use of LLMs with test-driven development (TDD) to generate code~\cite{Piya2023,Mock2025,Pancher2025}. Since TDD consists of incrementally creating test code before production code, which is only created to make the tests pass, the test code could be provided to the LLM to help the generation of the code and to verify the generated code. These results suggest that the work of software developers could become more focused on testing the code which is automatically generated. Another phenomenon pointing towards this direction is the term ``vibe coding'', coined by Andrej Karpathy, which has lately received a lot of attention among professionals~\cite{VibeCoding}. It represents a development workflow based on a conversational process between the developer and a chatbot, which takes advantage of the conversational interface common to notable tools based on LLMs, such as ChatGPT. In this workflow, developers spend a lot of their time inspecting the code generated by the chatbot rather than creating it themselves. 

In summary, the employment of GenAI to support software engineering will probably profoundly change software engineering practice, including the work of the professionals involved. More specifically, it is possible to software developers become more testers of the code generated by LLMs. These changes will prompt identity work by these professionals but, rather than the same for all individuals, it will be dependent on the role they assume in the development process. How this identity work will take place will have profound implications not only at individual level but also at organizational and even societal level. For example, role ambiguity and role conflict, i.e., workers struggling to balance assignments incompatible with their job expectations, have been associated with the productivity of development teams~\cite{Sun2018}.




\section{A Research Agenda}
\label{sec:agenda}

\begin{figure}[h]
  \centering
  \includegraphics[width=0.8\linewidth]{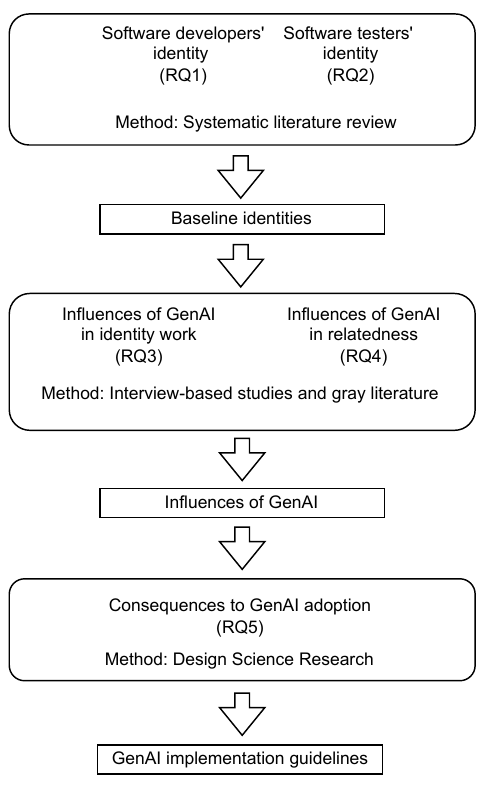}
  \caption{A framework to guide future research.}
  \label{fig:framework}
  \Description{A schematic summary of the research questions and related methods.}
\end{figure}

Based on the discussion above, there is a need to research the implications of the adoption of GenAI in software engineering to the work identities of the diverse roles involved. In particular, the tension between developers and testers, with the latter considered ``second-class citizens'', and a possible shift of the development to testing generated code highlight possible tensions among these roles and their identities. We are not considering the requirements engineer role given that previous research has shown that the prevalence of this role is limited~\cite{Herrmann2013}, being limited to large or very large companies~\cite{Wang2018}, and its responsibilities are generally delegated to other roles, such as product managers~\cite{Wang2020}. To tackle this research problem, we devised a set of research questions and the methods to be used to answer them as summarized in Fig.~\ref{fig:framework}.

The first step is to understand how different roles perceive themselves, i.e., their professional identities, and the differences among these perceptions. This goal is represented by the following research questions:\\

\noindent\textbf{RQ1: How do software developers perceive their own professional identity?}\\ \\
\textbf{RQ2: How do software testers perceive their own professional identity and how does it differ from software developers?}\\

To answer these research questions, we plan to perform a systematic literature review (SLR) on professional identity and related aspects of software developers and testers given the existence of several studies focused on this topic. Relying on existent studies also mitigates the issue of findings developers, testers and teams who have not adopted GenAI, which could be difficult given its pervasiveness.

The results obtained for RQ1 and RQ2 will serve as a baseline to understand how these professionals will adapt their identities, i.e., their identity work, with the adoption of GenAI for their work. In particular, it is important to identify potential source of conflicts triggering identity work and how they diverge for the different roles. To represent this goal, we propose the following research question:\\

\noindent\textbf{RQ3: How does the adoption of GenAI trigger identity work on different roles of software engineering, namely developers and testers?}\\

A relevant question within the context of the adoption of GenAI is how it will affect the way professionals perceive their peers in particular those with a different role in the process. This concern is represented by the following research question:\\

\noindent\textbf{RQ4: How does the adoption of GenAI influence the relatedness among different roles in software development?}\\

To answer RQ3 and RQ4, we plan to conduct qualitative studies based on diverse data sources. First, we foresee an interview-based study employing war stories, ``a form of qualitative data that capture informants' specific accounts of surmounting great challenges'' ~\cite{Lutters2007}. In such a way, we could identify events of identity work of developers and testers triggered by the adoption of GenAI. Second, to obtain a broader perspective, we plan to conduct a study based on practitioners-produced literature or gray literature~\cite{Garousi2019}.

Based on the results of previous research questions, we could explore the potential implications of the adoption of GenAI for software engineering, not only by investigating the state-of-practice but also by proposing new approaches, such as practices and process. These approaches could reduce tensions among team members and, consequently, facilitate the adoption of GenAI, leading to better results both in technical terms but also in human terms. As a provisional research question for this aspect, we propose the following:\\

\noindent\textbf{RQ5: How should GenAI be implemented to account for the identity work prompted by its adoption in software development teams?}\\

To achieve this goal, we intend to employ a Design Science Research (DSR) approach. DSR aims to produce prescriptive knowledge through the construction of socio-technical artifacts, which could be constructs, i.e. vocabulary and symbols, models, i.e., abstractions and representations, methods, i.e., algorithms and practices, or instantiations (implemented and prototype systems)~\cite{Hevner2004,Hevner2013,Engstrom2020}. More specifically, we expect to propose new constructs, models and probably methods, potentially reformulating roles in software teams combining the disruption of GenAI with the identity work triggered by it. These proposals could consist on the adaptation of these roles, with different responsibilities, or even new roles, and how these roles could interact in a team.  

We foresee several potential implications of the results to be obtained by this research agenda. First, it is known that professional identity is more adaptable and mutable in the beginning of a career~\cite{Ibarra1999} and education and training is an intensive period of identity work~\cite{Pratt2006}. These observations lead to the need for adapting the education of future software engineers and also developing training and re-skilling techniques for existent professionals to prepare them for a new context with reformulated roles. Some evidence already points out to concerns of some students on what they will become and if they will even have jobs, being replaced by GenAI~\cite{Melegati2024}. At team and organizational levels, the results of this agenda might inform the organization of development teams regarding their roles and related tasks and also hiring processes.

\section{Related work}
\label{sec:related_work}

In this section, we present a brief overview of the literature that will inform the execution of the research agenda, including seminal work on identity work. It also presents previous studies on human aspects of software engineering focusing on the implications of GenAI. 

Caza et al.~\cite{Caza2018} performed a review of the literature on identity work. They identified four theoretical approaches (social identity theory, critical theory, identity theory and narrative theory) that are mainly used to explain ``how, when, and why'' individuals engage in identity work and combined them to propose a holistic view of the phenomenon. The authors synthesize a comprehensive definition of identity work as ``cognitive, discursive, physical, and behavioral activities'' performed by individuals to form, repair, maintain, strength, revise or reject ``collective, role, and personal self-meanings''. These results indicate that the research to be conducted will need to combine different theoretical perspectives and research methods.

Regarding professional identities, by analyzing professionals transitioning to more senior roles, Ibarra~\cite{Ibarra1999} observed that the adaptation involved three basic tasks: the observation of role models to identify potential identities, the experimentation with possible identities, and the evaluation of the results of this experimentation through internal understanding and external feedback. The transition to a new role can be considered as a tension between the work identity and the work to be performed. In this sense, we can relate this result with those obtained by Pratt et al.~\cite{Pratt2006} in a longitudinal study with medical residents of three different areas: surgery, radiology and primary care. Based on that, the authors proposed a theory of identity construction consisted of two learning cycles, one about the work itself and the other about professional identity. Both cycles start with the work and assessments of work-identity integrity. In the first cycle, the identification of violations to work-identity leads to learning about the work which will be socially validated based on feedback. Some violations might trigger the second cycle in which professionals performed identity work by customizing and adapting their identities which was again validated, for example, based on role models. These results can provide theoretical lenses to investigate the identity work of software professionals triggered by the adoption of GenAI. However it would need to consider some concepts which might not be available given the particularities of the problem at hand. For example, given the novelty of the tools, there might be a lack of role models to these professionals.

In software engineering literature, there are already some studies that could support our studies or even present initial results related to our research agenda. Rauf and Sharp~\cite{Rauf2025} proposed a view of machines as team members rather than simply tools, they called Human-machine teaming (HMT) as a framework for researching AI for software engineering. Sghaier et al.~\cite{Sghaier2024} proposed SEWELL-CARE, an assessement framework for AI-driven software engineering tasks, considering technical, psychological and social aspects. As possible outcomes, the framework indicates work dynamics, including task conflict and role ambiguity. 
To investigate a hypotheses that the use of AI-based tools might influence the self-perceptions of software developers, Melegati et al.~\cite{Melegati2024} conducted a qualitative experiment with students, asking their perception of the role. The answers were labeled with the following themes: developer as a user of tools, developer as a builder, developer as an information seeker, and developer as a team player.

Schmitt et al.~\cite{Schmitt2025} conducted an interview-based study with 11 engineers of a large software organization regarding their identity work after the adoption of GenAI in the company. They observed that junior and senior software engineers felt that GenAI impact differently their needs for competence, autonomy and relatedness to the other engineers. For each aspect and level of experience, the authors identified ways that the engineers dealt with the impact of GenAI in their work, i.e., identity work. For example, regarding autonomy, junior developers saw a threat to their ownership and agency while senior developers felt a threat to the responsibility and control over the produced code. However, the authors did not evaluate the influence of the engineer role in their identity work triggered by the adoption of GenAI. Further studies are also needed to improve the generalizability of the results.

\section{Conclusions}
\label{sec:conclusions}

GenAI represents a possible disruption in the way software is developed, once its limitations, such as lack of determinism and explainability, are mitigated. A promising approach in this regard is the use of automatic tests to verify the generated outputs. Such an approach would mean that developers would become testers of the code produced by GenAI. However, this change might trigger identity work of software professionals. In this paper, we presented this research vision and how we intend to undertake research on this topic. We also discuss the potential implications of the results of such research.

\bibliographystyle{ACM-Reference-Format}
\bibliography{refs}



\end{document}